# Bitcoin's future carbon footprint

Shize Qin[1], Lena Klaaßen[2,4], Ulrich Gallersdörfer[3], Christian Stoll[4,5,*], Da Zhang[1,6,*]

[1] Institute of Energy, Environment, and Economy, Tsinghua University, Beijing 100084, China

[2] TUM School of Management, Technical University of Munich, Munich, Germany

[3] TUM Software Engineering for Business Information Systems, Department of Informatics, Technical University of Munich, Munich, Germany

[4] MIT Center for Energy and Environmental Policy Research, Massachusetts Institute of Technology, Cambridge, MA, USA

[5] TUM Center for Energy Markets, TUM School of Management, Technical University of Munich, Munich, Germany

[6] MIT Joint Program on the Science and Policy of Global Change, Cambridge, MA 02139, USA.

*Correspondence to Christian Stoll (cstoll@mit.edu) and Da Zhang (zhangda@tsinghua.edu.cn)

## Summary

The carbon footprint of Bitcoin has drawn wide attention, but Bitcoin's long-term impact on the climate remains uncertain. Here we present a framework to overcome uncertainties in previous estimates and project Bitcoin's electricity consumption and carbon footprint in the long term. If we assume Bitcoin's market capitalization grows in line with the one of gold, we find that the annual electricity consumption of Bitcoin may increase from 60 to 400 TWh between 2020 and 2100. The future carbon footprint of Bitcoin strongly depends on the decarbonization pathway of the electricity sector. If the electricity sector achieves carbon neutrality by 2050, Bitcoin's carbon footprint has peaked already. However, in the business-as-usual scenario, emissions sum up to 2 gigatons until 2100, an amount comparable to 7% of global emissions in 2019. The Bitcoin price spike at the end of 2020 shows, however, that progressive development of market capitalization could yield an electricity consumption of more than 100 TWh already in 2021, and lead to cumulative emissions of over 5 gigatons by 2100. Therefore, we also discuss policy instruments to reduce Bitcoin's future carbon footprint.

## Keywords

Bitcoin, gold, market capitalization, electricity consumption, carbon footprint, sensitivity analysis, climate goal, policy implication

# Introduction

The Bitcoin network uses a computationally-intensive process called "mining" to ensure the integrity of the system. The associated electricity consumption and carbon footprint of Bitcoin mining have received attention from researchers and policymakers alike.[1-3] Previous research shows that the electricity input to generate a $1 market value by Bitcoin mining (17 MJ) is higher than that of precious metal mining (5, 7, and 9 MJ for gold, platinum, and rare earth oxides, respectively)[1]. In total, the annual electricity consumption of Bitcoin mining (46 TWh in 2018) reaches the level of Portugal, and annual $CO_2$ emissions (22 megatons in 2018) match the level of a midsize city in the U.S.[2]

As Bitcoin becomes a more salient element of the global financial system, its growing carbon footprint has raised concerns about its role in climate change.[4-7] However, predicting Bitcoin mining's future electricity consumption and carbon footprint is associated with two key challenges. Predicting the future electricity efficiency of mining facilities, as well as predicting the future hash rate of the network, add significant uncertainty to estimates, given the high dynamics in the technological advancement of mining hardware and the price volatility of Bitcoin.[7-9]

Here we propose an alternative approach to project Bitcoin's electricity consumption and carbon footprint in the long term while avoiding the uncertainties discussed above. We use insight from the gold market as a proxy to estimate Bitcoin's future market capitalization. Bitcoin has been frequently compared to gold and other precious metals that may hedge the risk of severe losses during adverse economic conditions. Both Bitcoin and gold have been referred to as "safe-haven assets" due to their decentralized nature and inherent scarcity from a limited supply.[10-12] Taking the COVID-19 pandemic shock as an example, the year-to-date (as of November 30th, 2020) return of Bitcoin (190%) and gold (19%) outperformed the S&P 500 Index (13%).[13]

In addition to the insights from the gold market, the framework we present builds on the competitive market feature of the Bitcoin mining industry. The Bitcoin mining industry has no entry restrictions and limited economies of scale given constraints on the expansion of stable and low-cost electricity supply. Therefore, economic theory suggests that the revenues from mining should equal the total cost of mining in the long run. On the revenue side, the total revenue of Bitcoin mining in a specific year can be derived from the value of the Bitcoin mined plus fees paid for on-chain transactions. In our base case, we assume that Bitcoin's market capitalization grows with the average growth rate of gold's market capitalization over the last three decades (6%).[14-16] On the cost side, we conclude that electricity cost has been taking a relatively stable share of total mining costs (see Results). Based on the total electricity cost, we estimate the electricity consumption for Bitcoin mining by assuming a specific electricity price. We then utilize tailored emission factors – weighted by the geographical distribution of the hash rate – accounting for different decarbonization pathways of the electricity sector (see Methods).

We find that the electricity consumption of Bitcoin mining reaches about 400 TWh by 2100, about 2% of the current world's electricity consumption, as shown in Figure 1. The presented figures might be even higher if Bitcoin maintains its share compared to gold as of December 2020 (4.4%). Thus, the expected electricity consumption could already amount to 100 TWh in 2021. Bitcoin's carbon footprint, however, highly depends on the decarbonization rate of the world electricity sector. In a business-as-usual (BAU) scenario, cumulative $CO_2$ emissions reach 2 gigatons by 2100, about 7% of the world's total emissions in 2019 (~33 gigatons).[17] Under scenarios with higher decarbonization rates, total emissions would remain around 200 megatons – a non-negligible number but not decisive to limit global warming. Still, increased electricity consumption due to a strong performance in market capitalization of Bitcoin over the next decade would increase these estimates disproportionally leading to cumulative $CO_2$ emissions of even over 5 gigatons under a BAU scenario.

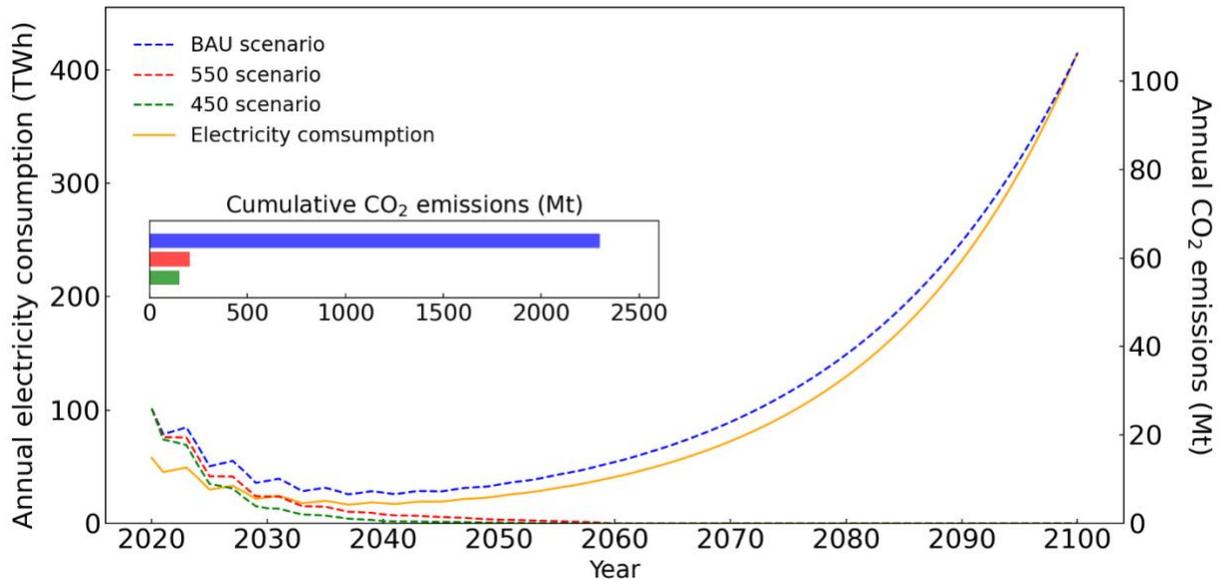

Fig. 1 | Annual electricity consumption and $CO_2$ emissions of Bitcoin mining through 2100. The business-as-usual (BAU) scenario assumes that the annual reduction rate of $CO_2$ emissions intensity of the world's electricity sector remains constant at 0.7%; the 450 scenario results in approximately 2°C global warming; the 550 scenario implies approximately 3°C global warming (for further details, see Carbon footprint of future Bitcoin mining below).

## Results

**Share of electricity cost in the total cost of Bitcoin mining.**

To estimate the future share of electricity cost in the total cost of Bitcoin mining, we analyze the historical development of this ratio and compare it with other computing processes that have similar characteristics (e.g., CPU computing or GPU computing). The trends in Figure 2a show that the share of electricity in total costs oscillates within a certain range. More extended time series of energy cost share in other energy-intensive industrial processes shown in Figure 2b support this observation. Therefore, we assume a constant share of electricity cost in total mining costs and can derive the future electricity cost from projected revenues of Bitcoin mining.

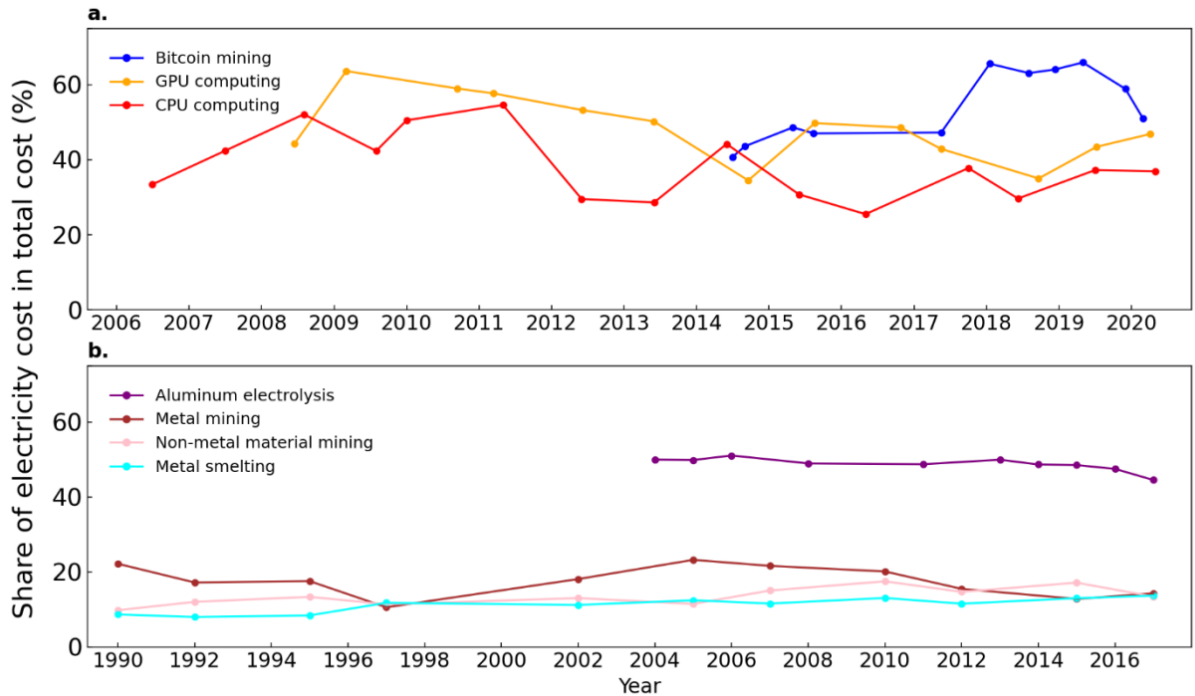

Fig. 2 | Share of electricity cost in total costs for Bitcoin mining and other energy-intensive processes.

**Electricity consumption of future Bitcoin mining.**

Today, the dominant source of Bitcoin mining revenue originates from block rewards (94% on average from January 1st, 2020 to November 30th, 2020).[18] As the block reward halves after every 210,000 blocks mined (approximately every four years; until zero by 2140), transaction fees for on-chain transactions will dominate the mining revenue in the long run. Therefore, we expect on-chain transaction fees to increase as Bitcoin's market capitalization grows to maintain an incentive for miners to participate in the network. In Bitcoin's current incentive structure, rising transaction fees are inevitable to secure the network, but they will also further fuel the energy-intensive mining process.

We find that the electricity consumption of Bitcoin mining will increase exponentially as transaction fees start to dominate mining revenue. Figure 3 shows the trend of mining revenue and associated electricity consumption until the end of this century. Over the next decades, we expect electricity consumption to experience a phase of decline due to shrinking block rewards. Electricity consumption driven by block rewards decreases to less than 1 TWh by 2053 compared to 54 TWh in 2020 and reaches zero by 2140. The trend changes around 2040 when electricity consumption returns to growth fueled by revenue from on-chain transaction fees, which grow in line with Bitcoin's market capitalization. By 2100, the annual electricity consumption reaches about 0.4 PWh, and continue to increase exponentially to 4 PWh by 2140; about one-fifth of the current world's electricity consumption.

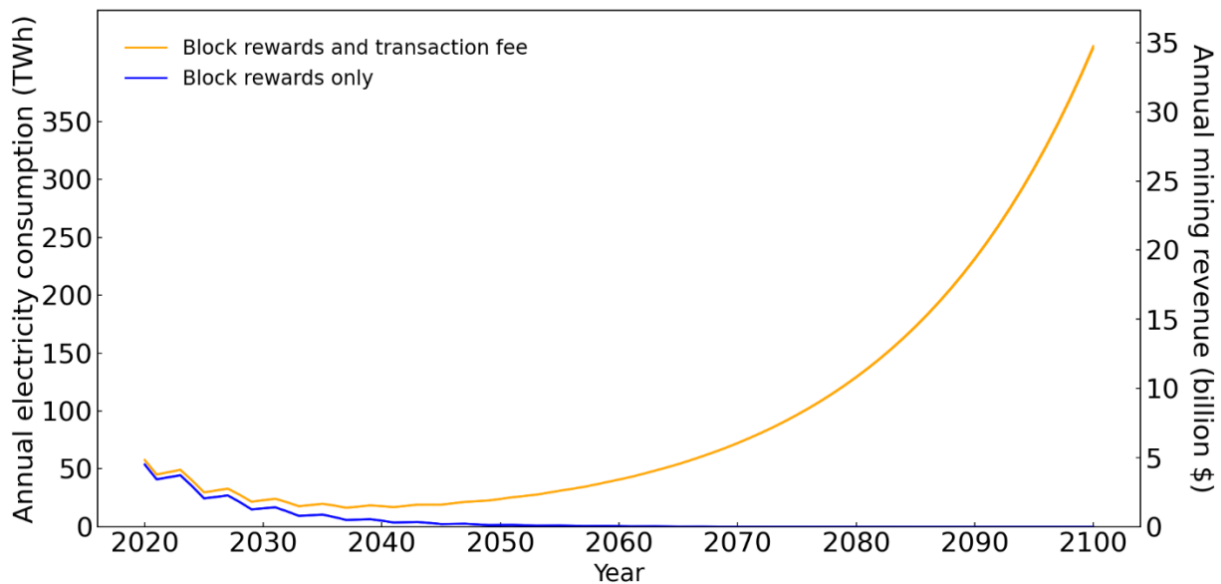

Fig. 3 | Mining revenue from block rewards and transaction fees and revenue from block rewards only, and their associated electricity consumption.

**Carbon footprint of future Bitcoin mining.**

$CO_2$ emission factors of the electricity used for mining are needed to translate the electricity consumption into carbon emissions. Since the carbon intensity of electricity generation varies greatly globally, the geographic distribution of mining activities is key to calculate the average $CO_2$ emission factor of the consumed electricity. However, the most comprehensive publicly available cryptocurrency mining map published by the Cambridge Center for Alternative Finance[19] only covers 35% of the Bitcoin network's hash rate (as of September 2020). Therefore, we collect independent information on the geographic distribution of mining activities for our estimate from mining pools. We collect the hash rates and geographical distribution of all major mining pools (market share of > 10%), which together account for 70% of the total network hash rates (from January 2020 to September 2020). Figure 4a shows the network hash rates distribution as of September 2020.

Since about 75% of the network hash rate originates from China, we further analyze the hash rate distribution among provinces within China. We find mining activities concentrated in regions with cheap electricity, such as Sichuan and Inner Mongolia. Mining hotspots outside China include Kazakhstan and Russia. Figure 4b shows the $CO_2$ emissions intensity of electricity generation globally,[20] and by region for China.[21-22] Based on these two data maps, we find the weighted average $CO_2$ emission factor of the electricity used for Bitcoin mining to be 0.46 kg/kWh as of 2020. The total $CO_2$ emissions we obtain for 2020 (24 megatons) is in line with previous research results.[2]

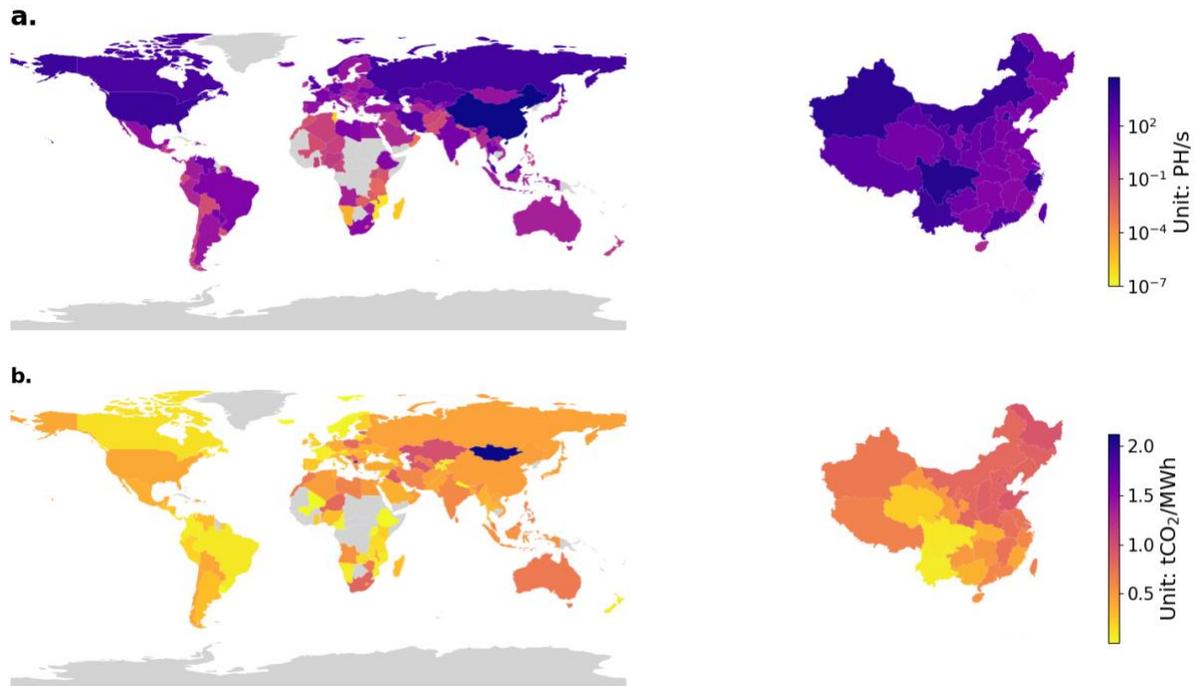

Fig. 4 | Hash rates distribution and emission factors of the electricity sector for world regions and Chinese provinces.

The future carbon footprint of Bitcoin mining strongly depends on the decarbonization pathway of the electricity sector. Therefore, we plot future carbon emissions from Bitcoin mining under different decarbonization pathways. In Figure 1, we show the business-as-usual (BAU) scenario, which assumes that the annual reduction rate of $CO_2$ emissions intensity of the world electricity sector remains constant at 0.7%.[23] The 450 and 550 scenarios use the average projected emissions intensity from selected integrated assessment models included in the IPCC AR5 Scenario Database (EMF27-450-Conv and EMF27-550-Conv).[24] These scenarios represent decarbonization pathways to stabilize $CO_2$ concentrations at 450 and 550 ppm, respectively by 2100. The 550 scenario results in a global temperature increase of about 3ºC, while the 450 scenario implies 2ºC warming by 2100.

Under the BAU scenario, carbon emissions increase in line with electricity consumption. The cumulative emissions until 2100 (~2 gigatons $CO_2$) equate to 7% of the world's total emissions in 2018 (~33 gigatons).[17] However, under the 450 and 550 scenarios with much faster decarbonization rates, $CO_2$ emissions from Bitcoin mining has peaked already around 2020 and gradually decline to zero by mid-century. This would lead to cumulative $CO_2$ emissions of around 200 megatons – a non-negligible number but not decisive for climate goals (limiting global warming to below 2°C allows for approximately 1,000 gigatons $CO_2$ emissions from now on)[25].

**Sensitivity analysis.**

We are aware that assumptions for some key parameters used in our estimation dictate the shape of our projected $CO_2$ emissions trajectories and Bitcoin's cumulative $CO_2$ emissions. These key parameters include the share of electricity in total mining cost, the annual growth rate of Bitcoin's market capitalization, and the annual decarbonization rate of the world electricity sector.

In the light of the price spike at the end of 2020, our assumption that Bitcoin maintains the same ratio to the market capitalization of gold in 2020 (1.9% from January 2020 to November 2020) seems rather conservative. In case, Bitcoin stabilizes its current share as of December 2020 (4.4%), the future electricity consumption and carbon emissions would already double compared to our base case. Thus, the electricity consumption in 2021 amounts to 100 TWh which compares to half of the world data center's energy use[26] and leads to cumulative emission of over 5 gigatons $CO_2$ emissions under a BAU scenario. This demonstrates the upside potential in case of a continued strong market capitalization performance.

To further facilitate the discussion on uncertainty, we normalize the initial emissions intensity of the world electricity sector in 2020 as one and approximate the decarbonization pathway of emissions intensity using a linear decarbonization rate. For instance, we use an annual decarbonization rate of 3% in the 550 scenario, which means the carbon intensity of the world electricity sector will decrease linearly and reach zero in about 33 years from 2020.

To demonstrate the sensitivity of our estimates relative to the above parameters, we first show in Figure 5 $CO_2$ emissions trajectories and cumulative $CO_2$ emissions until the world electricity sector achieves carbon neutrality under different parameter choices. We pick 0.5–0.7 as parameters for the share of electricity in total mining costs, 2–10% for the annual growth rate of Bitcoin's market capitalization, and 1–5% for the annual decarbonization rate of the world electricity sector. Changes in the annual decarbonization rate of the world electricity sector show a large impact on the emissions trajectory, while the share of electricity in total mining costs has a relatively small impact.

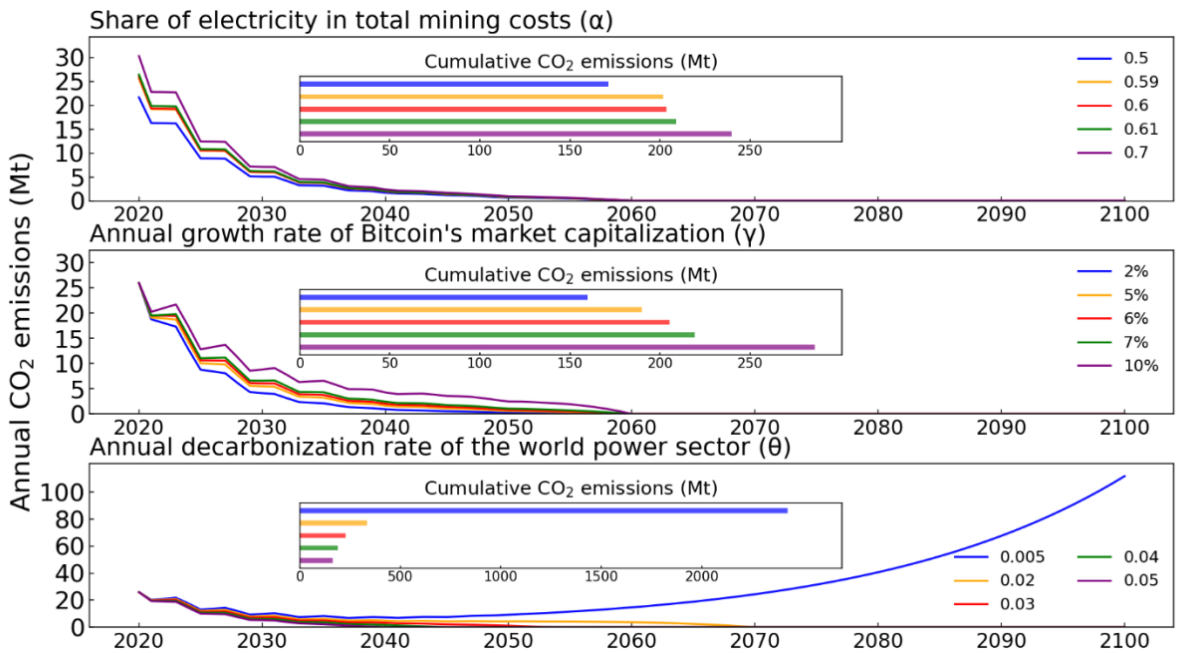

Fig. 5 | $CO_2$ emissions trajectories and cumulative $CO_2$ emissions of Bitcoin mining starting from 2021 under different parameter choices. We use 0.6, 6%, and 0.03 for the share of electricity in

total mining costs (α), annual growth rate of Bitcoin's market capitalization (γ), and annual decarbonization rate of the world power sector (θ), respectively, in our base case scenario.

Since all the three parameters are within the range between zero and one, we also derive the partial derivative of the cumulative $CO_2$ emissions relative to a percentage point change in each of the three parameters at the values for our main results; see details in the Methods section. Consistent with the finding above, the partial derivative for the annual decarbonization rate of the world electricity sector has the largest partial derivative, followed by the annual growth rate of Bitcoin's market capitalization. The share of electricity in total mining costs has the smallest partial derivative.

## Discussion

Our study provides a framework to analyze Bitcoin's future electricity consumption and carbon footprint. We find Bitcoin's annual electricity consumption to be 0.4 PWh and cumulative $CO_2$ emissions of 2 gigatons by 2100, respectively if the decarbonization of the world's electricity sector continues at the current speed. However, with much faster decarbonization that achieves carbon neutrality around the mid of the century (450 and 550 scenarios), Bitcoin's cumulative $CO_2$ emissions will not be decisive for limiting global warming to below 2°C. Besides the decarbonization rate of the world electricity sector, we show how the share of electricity in total mining cost for Bitcoin mining and the annual growth rate of Bitcoin's market capitalization shape Bitcoin's future carbon footprint. As a consequence, the presented results might be considerably higher if the share of Bitcoin compared to gold stabilizes as of December 2020. In this case, we can already expect an electricity consumption equal to half of the world data center's energy use by this year and cumulative $CO_2$ emissions of over 5 gigatons by 2100.

The proposed framework is not limited to Bitcoin but can also be applied to estimate the energy consumption and carbon footprint of other emerging technologies. This includes, for instance, additional cryptocurrencies[27], data centers, and communication infrastructures, given there is adequate information on the future size of the industry, and the energy share in total production costs.

Our analysis also delivers implications for policymakers that ponder on reducing the carbon footprint of Bitcoin mining. We focus the discussion on potential supply-side policy instruments that could reduce the share of electricity in total mining cost, or enhance the annual decarbonization rate of the world electricity system. First, any policies that can effectively raise the price of electricity used for Bitcoin mining will directly reduce the quantity of electricity consumption for creating a unit value of Bitcoin. There are also significant indirect effects of a higher electricity price, which includes the encouraged directed technical innovation that will trigger a reduction of the share of electricity in total mining cost.[28] Second, economy-wide carbon pricing entails a dual effect on reducing Bitcoin's carbon footprint by increasing the price of electricity and incentivizing the decarbonization of the world electricity system. Third, R&D subsidies may promote directed technical innovation, such as developing more energy-efficient integrated circuits (IC). More energy-efficient ICs as a result of directed technical innovation may lower the total electricity demand by reducing the share of electricity cost in total cost. Finally, we want to highlight that other regulative approaches, such as taxing mining activities or restricting the use of Bitcoin, may also lower Bitcoin's electricity and carbon footprint. Naturally, more significant factors are contributing to climate change. And yet, the carbon footprint of Bitcoin is large enough to make it worth discussing the possibility of regulating cryptocurrency mining in regions where electricity generation is especially carbon-intensive. However, implementing these regulations is controversial and remains an exciting direction for future research.

**Experimental Procedures**

**Overview of the analytical framework.**

We use the year 2020 as the starting point of the timeline $t$ in our analytical framework by setting $t = 0$ for 2020. We here estimate Bitcoin's current emissions in 2020, $E(0)$, and its cumulative emissions from 2021 ($t = 1$) through the year when the world electricity sector achieves carbon neutrality (we denote this year as $T$), $E^T$. We validate our analytical framework by comparing $E(0)$ to recent studies that estimate Bitcoin's annual emissions. Furthermore, we conduct a sensitivity analysis to demonstrate the impact of key parameters on $E^T$.

Bitcoin's cumulative emissions are the integration of its emissions in each year $t$, $E(t)$.

$$E^T = \int_1^T E(t)dt \tag{1}$$

Emissions in each year $t$ can be calculated by multiplying the electricity consumption, $ELE(t)$, by the average emissions intensity (weighted by the geographical distribution of hash rates) of the world electricity sector in that year, $\overline{EF}(t)$.

$$E(t) = ELE(t)\overline{EF}(t) \tag{2}$$

Applying the assumption that the mining revenue $R(t)$ is equal to mining cost $C(t)$ and electricity price will remain constant as $p^{ELE}$ ($0.05/kWh for 2019 based on our interviews with major miners), the electricity consumption can be estimated by dividing the expenses on electricity in mining activities by the electricity price. The expenses on electricity in mining activities can be

calculated using the share of electricity in total mining costs, $\alpha$, which is a key parameter that we will discuss later and include in the sensitivity analysis.

$$ELE(t) = \frac{\alpha C(t)}{p^{ELE}} = \frac{\alpha R(t)}{p^{ELE}} \tag{3}$$

We consider two revenue streams of the mining activity: block rewards and on-chain transaction fees, shown in the equation below. The first term on the right-hand side represents the revenue from block rewards. It can be calculated by multiplying the scheduled number of Bitcoins to be mined in year $t$, $q(t)$, by the price of Bitcoin in year $t$, which could be estimated using Bitcoin's market capitalization in year $t$, $V(t)$, and all the Bitcoins mined until year $t$, $Q(t)$. The second term represents the revenue from on-chain transaction fees, which is the product of Bitcoin's market capitalization in year $t$, $V(t)$, and a ratio of on-chain transaction fees to market capitalization, $\beta$, which is discussed later.

$$R(t) = \frac{V(t)}{Q(t)} q(t) + V(t)\beta \tag{4}$$

For our main results, we assume that Bitcoin's market capitalization will grow from $V(0)$ in year 0 at a constant rate, $\gamma$, which is a key parameter that we discuss later and include in the sensitivity analysis. Bitcoins mined by year $t$ is a pre-determined variable, which can be represented as the sum of Bitcoins mined by year 0, $Q(0)$, and all the Bitcoins mined from year 1 to year $t$, $\int_1^t q(t)dt$. Therefore, the mining revenue $R(t)$ can be rewritten as:

$$R(t) = \frac{V(0)(1+\gamma)^t}{Q(0) + \int_1^t q(t)dt} q(t) + V(0)(1+\gamma)^t \beta \tag{5}$$

Finally, we use different decarbonization scenarios of future average emissions intensity (weighted by the current geographical distribution of hash rates, assuming the distribution remains unchanged) of the world electricity sector in year $t$, $\overline{EF}(t)$. In the business-as-usual (BAU) decarbonization scenario, we assume the recent trend of annual reduction rate in the emissions intensity of the world electricity sector will continue. We adopt the annual reduction rate as 0.7% suggested by Knobloch et al.[23] for all the future years after 2020, and the world electricity sector cannot achieve carbon neutrality before the end of this century.

In two further decarbonization scenarios (450 scenario and 550 scenario), we consider faster decarbonization rates of the world electricity sector. We retrieve results for EMF27 analysis from integrated assessment models included in the IPCC AR5 Scenario Database (EMF27-450-Conv and EMF27-550-Conv[24] under the emission budget to stabilize $CO_2$ concentrations 450 and 550 ppm by 2100, respectively) with the trajectory of the world electricity sector's emissions intensity provided. These integrated assessment models are GCAM 3.0, MACLIM v1.1, MERGE_EMF27, POLES EMF27, REMIND 1.5, and WITCH_EMF27. We normalize the emissions intensity provided by these models in 2020 as one and calculate the average value of emissions intensity in later years for the 450 decarbonization scenario and 550 decarbonization scenario. Emissions intensity trajectories in all these three scenarios (relative to the emissions intensity in 2020) are shown in Figure S1. We use the emissions intensity trajectory of the 550 decarbonization scenario for our main results.

Combining all the equations above, Bitcoin's cumulative emissions for future years can be rewritten as:

$$E = \frac{\alpha}{p_{ELE}} \int_1^T V(0)(1+\gamma)^t \left[ \frac{q(t)}{Q(0) + \int_1^t q(t)dt} + \beta \right] \overline{EF}(t) dt \qquad (6)$$

**Annual growth rate of Bitcoin's market capitalization.**

We assume Bitcoin's market capitalization to grow at the average growth rate of gold's market capitalization over the last three decades (6%) and maintain its ratio to the gold's in the future (~2% on average from January 2020 to November 2020). We calculate the average market capitalization growth rate of gold using historical market capitalization data derived from historical gold price data from the Wind Financial Terminal[16] (converted to constant prices) and total historical supply of gold from the U.S. Geological Survey (USGS).[15] The average market capitalization growth rate of gold was 6% between 1994 (the earliest year with available data) and 2019, which we use as the value of $\gamma$ in our main results.

**Ratio of on-chain transaction fees to Bitcoin's market capitalization.**

To estimate Bitcoin's on-chain transaction fees given its capitalization, we again use gold as a reference. We notice that Bitcoin's on-chain transactions are comparable to the out-of-counter (OTC) trading of gold. For the OTC trading of gold, an independent organization, the London Bullion Market Association (LBMA), facilitates gold transactions outside centralized exchanges (e.g., Shanghai Gold Exchange). The LBMA sets the standard for the delivery and storage of gold and supports the clearing of the OTC trading. Similarly, core developers and miners of Bitcoin community perform comparable tasks of making Bitcoin improvement proposals (BIP) on major upgrades of the network and propagating transactions on Bitcoin's blockchain[29], respectively. Therefore, we adopt key parameters for the OTC trading of gold to infer the ratio of Bitcoin's on-chain transaction fees to its market capitalization, $\beta$. These parameters include the ratio of gold's total trading volume to its market capitalization ($\rho^{gold}$), the share of gold's OTC trading volume to its total trading volume ($\theta^{gold}$), and the transaction fee rate of OTC trading ($\varphi^{gold}$). In our main results, we assume the ratio of on-chain transaction fees to Bitcoin's market capitalization ($\beta$) equal to the ratio of OTC trading transaction fees to gold's market capitalization ($\beta^{gold}$).

$$\beta = \beta^{gold} = \rho^{gold}\theta^{gold}\varphi^{gold} \tag{7}$$

We estimate $\rho^{gold}$ by dividing the total trading volume of gold[30] by the total market capitalization of gold in 2017 (calculated in the above section). $\theta^{gold}$ is estimated by dividing the OTC trading volume by the total trading volume of gold in 2017. [30] We adopt the transaction fee rate of the gold trading agency service at Bank of China in 2020[31] as $\varphi^{gold}$. The values for $\rho^{gold}$, $\theta^{gold}$, and $\varphi^{gold}$ are 3.37, 0.58, and 0.0009, respectively, so the ratio of on-chain transaction fees to Bitcoin's market capitalization ($\beta$) is 0.0018. Applying this ratio in our framework gives the share of on-chain transaction fees in total mining revenue in 2020 as 6.8%, close to the observed share from January 2020 to November 2020 (6.4%).

**Share of electricity in total mining costs.**

We then analyze the evolution of the share of energy (i.e., electricity) in Bitcoin's total mining costs and compare it to other energy-intense processes such as CPU computing, GPU computing, aluminum electrolysis, metal mining (e.g., ferrous ore mining), non-metal mining (e.g., limestone mining), and metal smelting (e.g., iron smelting).

Bitcoin's share of electricity in total mining costs, $\alpha$, is estimated as follows (see details in the Supplementary Information). We collect key information for different generations of Application-specific integrated circuit (ASIC) mining hardware from a mainstream producer, Bitmain, based on information from its official website and social media accounts,[32-33] and calculate annualized electricity cost $c_i^{ele}$ and capital cost $c_i^{cap}$ per terahashes/second ($/(T/s)) of the mining using a specific type $i$ of Bitmain ASIC hardware.

$$\alpha_i = \frac{c_i^{ele}}{c_i^{ele} + c_i^{cap}} \tag{8}$$

The annualized electricity cost of mining per terahashes/second is calculated using the energy efficiency of mining hardware $i$, $HE_i$ (J/T), and electricity price, $p^{ele}$ ($/kWh). We adjust the electricity price to the release year of mining hardware $i$, $p_i^{ele}$ ($/kWh), using price indices provided by the OECD.[34]

$$c_i^{ele} = HE_i * p_i^{ele} * \frac{1}{1000} * 24 * 365 \tag{9}$$

The annualized capital cost of mining per hash rate is calculated using the release price of hardware $i$, $p_i$, total hash rates of hardware $i$, $HR_i$ (T/s), and market interest rate proxied by China's one-year loan interest rate $(r)$.[35] Here we assume the lifespan of mining hardware is 1.5 years based on our interview with some major miners.

$$c_i^{cap} = \frac{p_i}{HR_i * 1.5}(1 + r) \tag{10}$$

We adopt 0.6 for Bitcoin's share of electricity in total mining costs in our main results, $\alpha$, which is the average of calculated $\alpha_i$s from 2016 onwards.

We calculate the share of energy cost in total costs for six further energy-intensive processes to illustrate this value remains relatively stable over time. We first consider CPU computing and GPU computing per clockrate, which are similar energy-intensive computing processes like Bitcoin mining. Key performance parameters, such as clock rate (GHz) and thermal design power

(TDP) per clock rate (J/G), and release prices are collected from the official websites of Intel and Nvidia.[36-38] We assume the life span of CPU/GPU is four years.

We then consider aluminum electrolysis based on detailed production data in China. The annualized electricity cost of producing one ton of electrolytic aluminum in a specific year ($/t) is calculated by multiplying the average electricity intensity of the production process (kWh/t)[39] by the coal-fired on-grid electricity price in that year ($/kWh).[40-41] The annualized capital cost is the sum of annualized alumina cost, anodes cost, cathode cost, aluminum fluoride cost, cryolite cost, and labor cost for producing a ton of electrolytic aluminum ($/t), based on a third-party data aggregator.[42]

We also calculate the share of energy cost in total costs for metal mining, non-metal mining, and metal smelting sector in China using multiple Chinese input-output tables from 1990 to 2017.[43] The energy cost in a specific year is the sum of the value of energy inputs, including coal, oil, gas, and electricity, and the total cost is calculated by summing the value of all the inputs.

**Emissions intensity of current mining networks.**

We denote the emissions intensity of current mining networks, which is the average emissions intensity weighted by the geographical distribution of hash rates in 2020, as $\overline{EF}(0)$.

$$\overline{EF}(0) = \sum_r EF_r^{2020} * s_r^{2020} + \sum_s EF_s^{2020} * s_s^{2020} \qquad (11)$$

where $r$ represents different Chinese provinces, $s$ represents different world regions except for China, $EF_r^{2020}$ or $EF_s^{2020}$ is the emissions intensity of the electricity sector in region $r$ or $s$ in 2020, and $s_r^{2020}$ or $s_s^{2020}$ is the share in total network hash rates for region $r$ or $s$ in 2020. Since the most recent data for the emissions intensity of the electricity sector by region are available for

2017, we apply an annual reduction rate as 0.7% suggested by Knobloch et al.[23] to scale the emissions intensity in 2017 ($EF_r^{2020}$ or $EF_s^{2020}$) to acquire the emissions intensity in 2020.

We calculate the emissions intensity of the electricity sector for a world region $s$ other than China, $EF_s^{2017}$, using emissions from electricity and CHP production as well as electricity production data provided by the International Energy Agency.[44] To calculate $EF_r^{2017}$ for a Chinese province $r$, we use operating margin (OM) emission factors of the grid region $R$ that $r$ belongs to, $EF_{R(r)}^{2017,OM}$,[45] and the share of coal-fired electricity (the marginal electricity) in this grid region in 2017, $s_{R(r)}^{2017,coal}$.[46]

$$EF_r^{2017} = EF_{R(r)}^{2017,OM} * s_{R(r)}^{2017,coal} \qquad (12)$$

We validate the above estimation by showing the weighted average emissions intensity of China (0.64 kg/kWh) using $EF_r^{2017}$ and electricity generation of each province is close to IEA's emissions intensity for China (0.69 kg/kWh) in 2017.[47]

To estimate the share of hash rates in a specific region $j$ ($j = r, s$) in 2020, $s_j^{2020}$, we use the information about the total hash rates of five major mining pools (indexed by $i$, covering 70% of the network's hash rates) and hash rates geographic distribution information for each of these pools. We first estimate the share of each mining pool's total hash rates in the network's hash rates, $w_i^{2020}$, by dividing the total number of blocks generated by each mining pool by the total number of blocks generated by these five pools from September 2019 to September 2020.[48] If we have the share of hash rates in a specific region $j$ for a specific mining pool $i$, $s_{i,j}^{2020}$, the share of hash rates in a specific region $j$ can be estimated as below.

$$s_j^{2020} = \sum_i w_i^{2020} * s_{i,j}^{2020} \tag{13}$$

However, acquiring $s_{i,j}^{2020}$ is a major challenge noted by existing studies[1-3] as there is very limited public information. To estimate $s_{i,j}^{2020}$, we collect the total hash rates and hash rates by region within a certain period between 2019 and 2020 for each of these mining pools through a series of in-depth interviews. All the pools have provided aggregated hash rates for China, $HR_{i,China}^{2020}$, and for the rest of the world, $HR_{i,ROW}^{2020}$, but only a subset of pools has provided disaggregated hash rates for Chinese provinces or regions in the rest of the world. We use the geographic distribution information from pools that disaggregated hash rates are available to infer the disaggregated hash rates for pools without the information. For example, we index the pools that have provided disaggregated hash rates for Chinese provinces by $ic$, then the hash rates of a Chinese province $r$ for a pool $i$ without disaggregated hash rates for Chinese provinces can be estimated as follows.

$$HR_{i,r}^{2020} = HR_{i,China}^{2020} * \frac{\sum_{i \in ic} HR_{i,r}^{2020}}{\sum_{i \in ic} HR_{i,China}^{2020}} \tag{14}$$

Similarly, the hash rates of a region in the rest of the world $s$ for a pool $i$ without disaggregated hash rates for the rest of the world can be estimated. We then have a complete dataset of hash rates for all the Chinese provinces and regions in the rest of the world, $HR_{i,j}^{2020}$, so estimating $s_{i,j}^{2020}$ is straight-forward.

$$s_{i,j}^{2020} = \frac{HR_{i,j}^{2020}}{\sum_j HR_{i,j}^{2020}} \tag{15}$$

**Sensitivity analysis.**

To facilitate the sensitivity analysis, we use linear trends to fit emissions intensity trajectories, so emissions intensity beyond 2020, $EF(t)$, can be represented as the equation below, where $\theta$ is the annual decarbonization rate of the world electricity sector. For example, our main results use the 550 scenario, $\theta \approx 0.03$, which means the world electricity sector will become carbon neutral within roughly 33 years after 2020.

$$EF(t) = EF(0)(1 - \theta t) \tag{16}$$

Bitcoin's cumulative emissions for future years could be rewritten as:

$$E = \frac{\alpha}{p_{ELE}} \int_1^T V(0)(1+\gamma)^t \left[ \frac{q(t)}{Q(0) + \int_1^t q(t)dt} + \beta \right] EF(0)(1 - \theta t)dt \tag{17}$$

We are interested in changes in Bitcoin's cumulative emissions for future years ($E$) relative to changes in the share of electricity in total mining cost ($\alpha$), the annual growth rate of Bitcoin's market capitalization ($\gamma$), and the annual decarbonization rate of the world electricity sector ($\theta$). We derive these partial derivatives as below.

$$\begin{aligned}
\frac{\partial \log E}{\partial \alpha} &= \frac{1}{E} \frac{\partial E}{\partial \alpha} \\
&= \frac{1}{E} \frac{1}{p_{ELE}} \int_1^T V(0)(1+\gamma)^t \left[ \frac{q(t)}{Q(0) + \int_1^t q(t)dt} + \beta \right] EF(0)(1 - \theta t)dt \\
&= \frac{1}{\alpha}
\end{aligned} \tag{18}$$

$$\begin{aligned}
\frac{\partial \log E}{\partial \gamma} &= \frac{1}{E} \frac{\partial E}{\partial \gamma} \\
&= \frac{1}{E} \frac{1}{p_{ELE}} \int_1^T \alpha V(0)(1+\gamma)^{t-1} \left[ \frac{q(t)}{Q(0) + \int_1^t q(t)dt} + \beta \right] EF(0)(1 - \theta t) t \, dt
\end{aligned} \tag{19}$$

$$-\frac{\partial logE}{\partial \theta} = \frac{1}{E}\left(-\frac{\partial E}{\partial \theta}\right)$$
$$= \frac{1}{E}\frac{1}{p_{ELE}}\int_1^T \alpha V(0)(1+\gamma)^t \left[\frac{q(t)}{Q(0)+\int_1^t q(t)dt} + \beta\right] EF(0) \quad (20)$$

We first compare the magnitude of $-\frac{\partial logE}{\partial \theta}$ and $\frac{\partial logE}{\partial \gamma}$.

$$-\frac{\partial logE}{\partial \theta} - \frac{\partial logE}{\partial \gamma}$$
$$= \frac{1}{E}\frac{EF(0)}{p_{ELE}}\int_1^T \alpha V(0)(1+\gamma)^{t-1} \left[\frac{q(t)}{Q(0)+\int_1^t q(t)dt} + \beta\right](\gamma + \theta t)t dt > 0 \quad (21)$$

Therefore, a percentage point change in the annual decarbonization rate of the world electricity sector has a larger effect on Bitcoin's cumulative emissions than a percentage point change in the annual growth rate of Bitcoin's market capitalization. Plugging the numerical values of all the parameters used in our main results, we obtain $-\frac{\partial logE}{\partial \theta} \approx 29$ and $\frac{\partial logE}{\partial \gamma} = 9.0$.

We then compare the magnitude of $\frac{\partial logE}{\partial \gamma}$ and $\frac{\partial logE}{\partial \alpha}$.

$$\frac{\partial logE}{\partial \gamma} - \frac{\partial logE}{\partial \alpha}$$
$$= \frac{1}{E}\frac{1}{p_{ELE}}\int_1^T V(0)(1+\gamma)^{t-1}\left[\frac{q(t)}{Q(0)+\int_1^t q(t)dt}+\beta\right] EF(t)[\alpha t - (1+\gamma)]dt \quad (22)$$

The above equation has negative values when $t < \frac{1+\gamma}{\alpha}$ and turns positive when $t > \frac{1+\gamma}{\alpha}$. There is no close form solution for the condition under which $\frac{\partial logE}{\partial \gamma} - \frac{\partial logE}{\partial \alpha} > 0$, but plugging the numerical values of all the parameters used in our main results, we obtain $\frac{\partial logE}{\partial \gamma} = 9.0$ and $\frac{\partial logE}{\partial \alpha} = 1.7$. We have validated numerically that $\frac{\partial logE}{\partial \gamma}$ is larger than $\frac{\partial logE}{\partial \alpha}$ within the reasonable range for $\alpha$ ($0.3 \leq \alpha \leq 1$) and $\gamma$ ($0.01 \leq \gamma \leq 0.2$). Therefore, a percentage point change in the annual growth rate of Bitcoin's market capitalization has a larger effect on Bitcoin's cumulative emissions than a percentage point change in the share of electricity in total mining cost.

## Acknowledgments

This study has been supported by the National Natural Science Foundation of China (71974109 and 71690244). We thank Ziheng Zhu and Xiaomo Wang for excellent research assistance.

## Author Contribution

C.S. and D.Z. conceived the study. All authors contributed to the design of the study. S.Q., C.S. and D.Z. aggregated and analyzed the data. S.Q., C.S. and D.Z. drafted the manuscript. U.G., L.K. and C.S. reviewed several drafts, made substantial revisions, and provided additions.

## Data and code availability

The codes of the sensitivity analysis are available in the online supplementary files. Additional scripts used for data analysis can be requested from the corresponding authors. The data supporting the findings of this study are available within the Article and Supplementary Information. The raw hash rate distribution data are available from the corresponding authors.

# Supplementary Information for

## "Bitcoin's future carbon footprint"

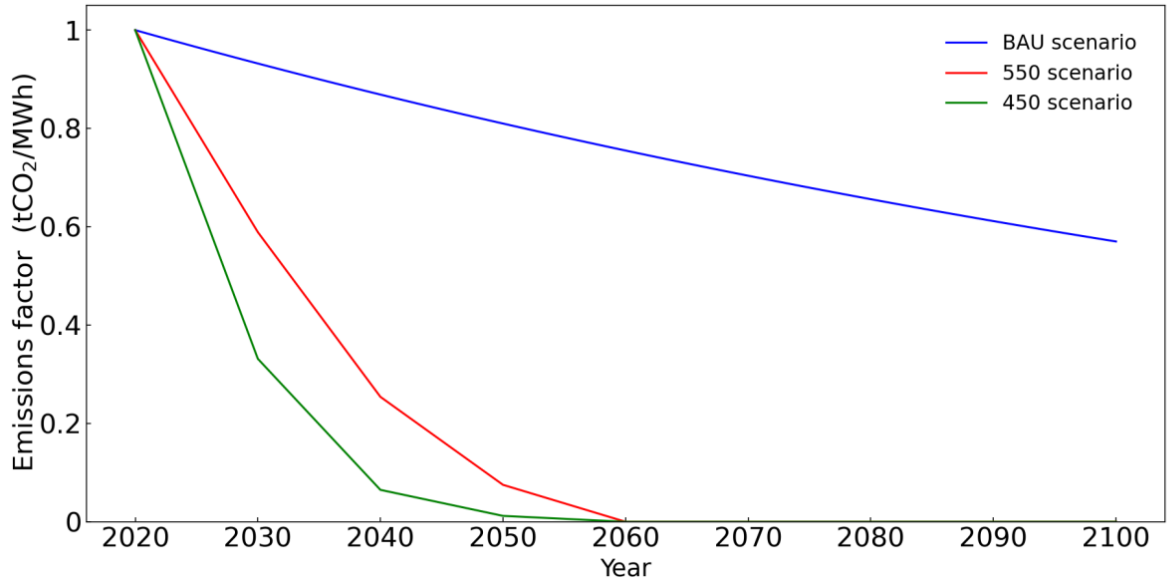

**Figure S1.** Emissions intensity trajectories of the world electricity sector in the BAU, 450, and 550 scenario (emissions intensity in 2020 is normalized as 1).

Supplementary information for supporting data and calculations (.xlsx file), script for sensitivity analysis (.py file) and data input for sensitivity analysis (.csv file) are submited in separate files upon submission.